\documentclass[12pt, fleqn, a4paper]{article} 
 \usepackage{epsfig, subcaption}
 \usepackage{hyperref}
 \hypersetup{pdfstartview = FitH,
             pdfborder    = {0 0 0.5},
             colorlinks   = false}
 \usepackage{color}
 \newcommand{\rmXA}      [2] {\sp{#2} {\rm #1}}
\newcommand{\mchi}           {m_{\chi}}
\newcommand{\mN}             {m_{\rm N}}
\newcommand{\mrN}            {m_{\rm r, N}}
\newcommand{\sigmaSI}        {\sigma_0^{\rm SI}}
\newcommand{\sigmaSD}        {\sigma_0^{\rm SD}}
\newcommand{\sigmapSI}       {\sigma_{\chi {\rm p}}^{\rm SI}}
\newcommand{\armp}           {a_{\rm p}}
\newcommand{\armn}           {a_{\rm n}}
\newcommand{\FSIQ}           {F_{\rm SI}^2(Q)}
\newcommand{\FSDQ}           {F_{\rm SD}^2(Q)}
\newcommand{\vchiLab}        {v _{\chi, {\rm Lab}}}
\newcommand{\thetaNRchi}     {\theta_{\rm N_R, \chi_{in}}}
 \newcommand{\rmF}           {\rmXA{F}  {19}}
 \newcommand{\rmAr}          {\rmXA{Ar} {40}}
 \newcommand{\rmGe}          {\rmXA{Ge} {73}}
 \newcommand{\rmXe}          {\rmXA{Xe}{129}}
 \newcommand{\rmW}           {\rmXA{W} {183}}
\newcommand{\OnlineSpectrum} [3] {
\href{http://www.tir.tw/phys/hep/dm/amidas-2d/amidas-2d.php%
      ?amidas_2D_function=NR_ang%
      &mode_NR=NR_ang_2D\%2B%
      &frame=geoG%
      &mode_animation=spectrum2D\%2B%
      &target=#1%
      &mchi=#2%
      &period=periodA%
      &Q_range=0_5_10_15_20%
      &event_No=500}
     {\fcolorbox {green} {white}
       {\begin{minipage} {17.2 cm}
         \begin{center}
          \vspace{0.1  cm}
           \includegraphics [width = 4.2 cm]
            {NR_ang_2D+-#1-\WIMPmass-geoG-0000_0050-0500-00000}%
           \includegraphics [width = 4.2 cm]
            {NR_ang_2D+-#1-\WIMPmass-geoG-0050_0100-0500-00000}%
           \includegraphics [width = 4.2 cm]
            {NR_ang_2D+-#1-\WIMPmass-geoG-0100_0150-0500-00000}%
           \includegraphics [width = 4.2 cm]
            {NR_ang_2D+-#1-\WIMPmass-geoG-0150_0200-0500-00000}%
          \\ \vspace{-0.1  cm}
          #3
          \\ \vspace{0.1  cm}
         \end{center}
        \end{minipage}}}
}
\newcommand{\InsertPlotNRang} [2] {
\begin{figure} [t!]
\begin{center}
 \OnlineSpectrum
  {F19}
  {#1}
  {$\rmF$}
 \\ \vspace{ 0.2 cm}
 \OnlineSpectrum
  {Ar40}
  {#1}
  {$\rmAr$}
 \\ \vspace{ 0.2 cm}
 \OnlineSpectrum
  {Ge73}
  {#1}
  {$\rmGe$}
 \\ \vspace{ 0.2 cm}
 \OnlineSpectrum
  {Xe129}
  {#1}
  {$\rmXe$}
 \\ \vspace{ 0.2 cm}
 \begin{subfigure} [c] {4.2 cm}%
 \caption{ 0 --  5 keV}
 \end{subfigure}
 \begin{subfigure} [c] {4.2 cm}%
 \caption{ 5 -- 10 keV}
 \end{subfigure}
 \begin{subfigure} [c] {4.2 cm}%
 \caption{10 -- 15 keV}
 \end{subfigure}
 \begin{subfigure} [c] {4.2 cm}%
 \caption{15 -- 20 keV}
 \end{subfigure}
 \\ \vspace{-0.6 cm}
\end{center}
\caption{
  #2
}
\label{fig:NR_ang_2D+-\WIMPmass-geoG-0500-00000}
\end{figure}
}
\begin{document}
\thispagestyle{empty}
\begin{flushright}
 August 2022
\end{flushright}
\begin{center}
{\Large\bf
 A Possibility of
 Determining the WIMP Mass by                          \\
 Using the Angular Recoil--Energy Spectra from         \\ \vspace{0.155 cm}
 Directional Direct Dark Matter Detection Experiments} \\
\vspace*{0.7 cm}
 {\sc Chung-Lin Shan}                                  \\
\vspace{0.5 cm}
 {\small\it
  Preparatory Office of
  the Supporting Center for
  Taiwan Independent Researchers                       \\ \vspace{0.05  cm}
  P.O.BOX 21 National Yang Ming Chiao Tung University,
  Hsinchu City 30099, Taiwan, R.O.C.}                  \\~\\~\\
 {\it E-mail:} {\tt clshan@tir.tw}
\end{center}
\vspace{2 cm}
\begin{abstract}

 In this article,
 as an extension of
 our study on
 the angular distribution of
 the recoil flux of
 WIMP--scattered target nuclei,
 we demonstrate
 a possibility of
 determining the mass of incident halo WIMPs
 by using or combining
 ``ridge--crater'' structures of
 the angular recoil--energy spectra
 with different target nuclei
 observed in directional direct Dark Matter detection experiments.
 Our simulation results show that,
 for a WIMP mass of
 only a few tens GeV,
 the stereoscopic angular recoil--flux distributions of
 both of light and heavy target nuclei
 would have a (longitudinally) ``ridge--like'' structure.
 However,
 once the WIMP mass is as heavy as a few hundreds GeV,
 the angular recoil--flux distributions of
 heavy target nuclei
 would in contrast show
 a (latitudinally) ``crater--like'' structure.

\end{abstract}
\clearpage
\section{Introduction}

 Direct Dark Matter (DM) detection experiments
 aiming to observe scattering signals of
 Weakly Interacting Massive Particles (WIMPs)
 off target nuclei
 would still be the most reliable experimental strategy
 for identifying Galactic DM particles
 and determining their properties
 \cite{SUSYDM96, Schumann19, Baudis20, Cooley21}.
 While most direct DM detection experiments
 measure only recoil energies
 deposited in underground detectors,
 the ``directional'' direct detection experiments
 could provide additional 3-dimensional information
 (recoil tracks and/or head--tail senses)
 of (elastic) WIMP--nucleus scattering events
 for discriminating WIMP signals
 from isotropic backgrounds
 and/or some incoming--direction--known astronomical events
 \cite{Ahlen09, Mayet16, Vahsen21}.

 In Refs.~\cite{DMDDD-3D-WIMP-N, DMDDD-NR},
 we have developed
 the double Monte Carlo scattering--by--scattering simulation package
 for the 3-dimensional elastic WIMP--nucleus scattering process
 and studied
 the angular distribution of
 the recoil flux of
 WIMP--scattered target nuclei
 in different celestial coordinate systems.
 Then,
 in Ref.~\cite{DMDDD-v_theta},
 we have shown
 the angular recoil--flux distributions of
 different target nuclei
 in several narrow recoil energy windows
 in the Equatorial and (geocentric) Galactic coordinate systems
 and observed
 the WIMP--mass and target--nucleus dependence of
 such angular recoil--energy spectra.
 In this article,
 we discuss therefore
 a possibility of
 determining the mass of incident halo WIMPs
 by using or combining
 the angular recoil--energy spectra
 observed (hopefully in the future)
 in directional direct DM detection experiments
 (with different target nuclei).

 In the next section,
 we will consider two WIMP masses
 and present the corresponding angular recoil--energy spectra
 of different target nuclei
 whose masses
 could basically cover
 the mass range of almost all nuclei
 used in direct DM detection experiments.
 Then
 we conclude in Sec.~3.

\section{Angular recoil--energy spectra}

 This work is based on
 our double Monte Carlo (MC) simulations
 for 3-D elastic WIMP--nucleus scattering
 described in detail in Refs.~\cite{DMDDD-3D-WIMP-N, DMDDD-NR}:
 we MC generate first a {\em 3-D velocity} of
 an incident WIMP
 in the {\em Galactic} coordinate system
 according to the theoretical isotropic Maxwellian velocity distribution,
 transform it to the laboratory coordinate system,
 and,
 in the laboratory
 (more precisely,
  the incoming--WIMP) coordinate system,
 we generate an {\em equivalent recoil angle} of
 a scattered target nucleus%
\footnote{
 The elevation of the nuclear recoil direction
 in the incoming--WIMP coordinate system,
 namely,
 the complementary angle of the recoil angle
 \cite{DMDDD-3D-WIMP-N}.
}
 and validate this candidate scattering event
 according to the criterion
 \cite{DMDDD-3D-WIMP-N}:
\begin{equation}
     f_{\rm N_R}(\vchiLab, \thetaNRchi)
  =  \frac{\vchiLab}{v_{\chi, {\rm cutoff}}}
     \bigg[\sigmaSI \FSIQ + \sigmaSD \FSDQ\bigg]
     \sin(2 \thetaNRchi)
\,.
\label{eqn:f_NR_thetaNRchi}
\end{equation}
 Here $\vchiLab$ and $\thetaNRchi$ are
 the transformed WIMP incident velocity
 in the laboratory coordinate system
 and the generated equivalent recoil angle of
 the scattered target nucleus,
 $v_{\chi, {\rm cutoff}}$
 is a cut--off velocity of incident halo WIMPs
 in the Equatorial/laboratory coordinate systems,
 which is set as 800 km/s
 in our simulations,
 $\sigma_0^{\rm (SI, SD)}$ are
 the spin--independent/dependent (SI/SD) total cross sections
 ignoring the nuclear form factor suppressions,
 and
 $F_{\rm (SI, SD)}(Q)$ indicate
 the elastic nuclear form factors
 corresponding to the SI/SD WIMP interactions,
 respectively.
 Note that
 the recoil energy of the scattered target nucleus
 appearing in Eq.~(\ref{eqn:f_NR_thetaNRchi})
 is now
 estimated by the equivalent recoil angle
 $\thetaNRchi$
 \cite{DMDDD-3D-WIMP-N}:
\begin{equation}
     Q(\thetaNRchi)
  =  \left[\left(\frac{2 \mrN^2}{\mN}\right) \vchiLab^2\right]
     \sin^2(\thetaNRchi)
\,,
\label{eqn:QQ_thetaNRchi}
\end{equation}
 where
\(
         \mrN
 \equiv  \mchi \mN / (\mchi + \mN)
\)
 is the reduced mass of
 the WIMP mass $\mchi$ and
 that of the target nucleus $\mN$.

 As in our earlier works
 \cite{DMDDD-NR,
       DMDDD-v_theta},
 in our simulations
 presented in this article,
 the SI (scalar) WIMP--nucleon cross section
 has been fixed as $\sigmapSI = 10^{-9}$ pb,
 while
 the SD (axial--vector) WIMP--proton/neutron couplings
 have been tuned as $\armp = 0.01$
 and $\armn = 0.7 \armp = 0.007$,
 respectively.
 Considering the exponential--like decreased
 scattering event rate
 with increasing the recoil energy,
 the simulation energy range
 has been commonly limited between 0 and 20 keV
 for all considered target nuclei
 and
 sliced into four 5-keV energy windows
 \cite{DMDDD-v_theta}.
 5,000 experiments
 with 500 accepted
 WIMP scattering events on average
 (Poisson--distributed)
 in one entire year
 in one experiment
 for one target nucleus
 have been simulated.

\subsection{A light WIMP mass of 20 GeV}

 In this subsection,
 we consider at first
 the case of a light WIMP mass of $\mchi = 20$ GeV.

 \def \WIMPmass    {0020}
 \InsertPlotNRang
  {20}
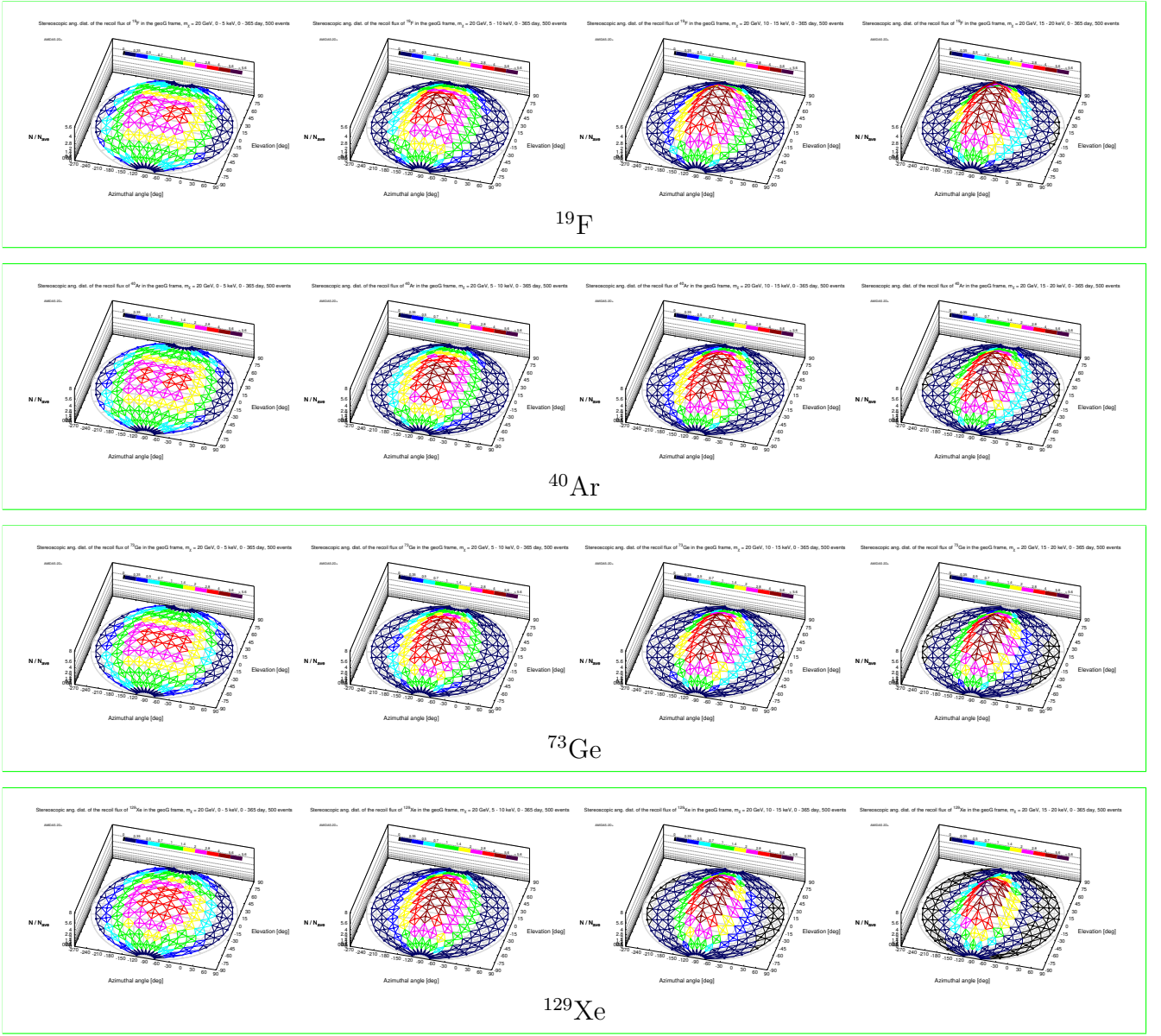
  {The stereoscopic angular recoil--flux distributions
   induced by 20-GeV WIMPs
   (in unit of their all--sky average values)
   observed in the {\em geocentric} Galactic coordinate system
   in the energy range between 0 and 20 keV
   \cite{DMDDD-v_theta}.
   Four nuclei:
   $\rmF$,
   $\rmAr$,
   $\rmGe$,
   and $\rmXe$
   have been considered.
   Note that
   the distribution centers
   have been shifted to 90$^{\circ}$W,
   so that
   the theoretical main direction of incident WIMPs
   \cite{DMDDD-3D-WIMP-N}:
   0.60$^{\circ}$S, 98.78$^{\circ}$W,
   is approximately at the center of our plots.
   See the text for further details.%
   }

 In Figs.~\ref{fig:NR_ang_2D+-0020-geoG-0500-00000},
 we show
 the {\em stereoscopic} angular distributions of
 the WIMP--induced nuclear recoil flux
 (in unit of their all--sky average values)
 observed in the {\em geocentric} Galactic coordinate system
 in four 5-keV recoil energy windows,
 respectively%
\footnote{
 Interested readers can click each row
 in Figs.~\ref{fig:NR_ang_2D+-0020-geoG-0500-00000}
 and \ref{fig:NR_ang_2D+-0200-geoG-0500-00000}
 to open the corresponding webpage of
 animated demonstrations
 (for more considered WIMP masses)
 \cite{AMIDAS-2D-web}.
}.
 Four nuclei
 used frequently
 in (directional) direct detection experiments:
 $\rmF$,
 $\rmAr$,
 $\rmGe$,
 and $\rmXe$
 have been considered
 as our targets.
 Note that
 500 accepted
 WIMP scattering events on average
 in ``each (5-keV)'' energy window/plot
 have been simulated
 and binned into 12 $\times$ 12 bins
 for the azimuthal angle and the elevation,
 respectively.
 Note also that
 the distribution centers
 have been shifted to 90$^{\circ}$W,
 so that
 the theoretical main direction of incident WIMPs
 (the opposite direction of the Solar Galactic movement
  toward the Cygnus constellation)
 \cite{DMDDD-3D-WIMP-N}:
 0.60$^{\circ}$S, 98.78$^{\circ}$W,
 is approximately at the center of our plots.

 With these stereoscopic demonstrations,
 one can clearly observe
 a (longitudinally) ``ridge--like'' structure of
 the angular recoil--flux distributions
 in the energy range between 5 and 20 keV:
 the higher the recoil energy (window),
 the more concentrated the angular recoil--flux distribution
 and (relatively) the higher the crest--line/peak.
 Meanwhile,
 Figs.~\ref{fig:NR_ang_2D+-0020-geoG-0500-00000}
 show that,
 the heavier the target nuclei,
 the more obviously and rapidly
 the ridge--like structure of
 the angular recoil--flux distributions
 as well as
 the concentration and the increase of
 the angular recoil--energy spectrum
 would be.

\subsection{A heavy WIMP mass of 200 GeV}

 As a comparison,
 in Figs.~\ref{fig:NR_ang_2D+-0200-geoG-0500-00000}
 the mass of incident WIMPs has been raised to $\mchi = 200$ GeV.

 \def \WIMPmass    {0200}
 \InsertPlotNRang
  {200}
  {As Figs.~\ref{fig:NR_ang_2D+-0020-geoG-0500-00000}:
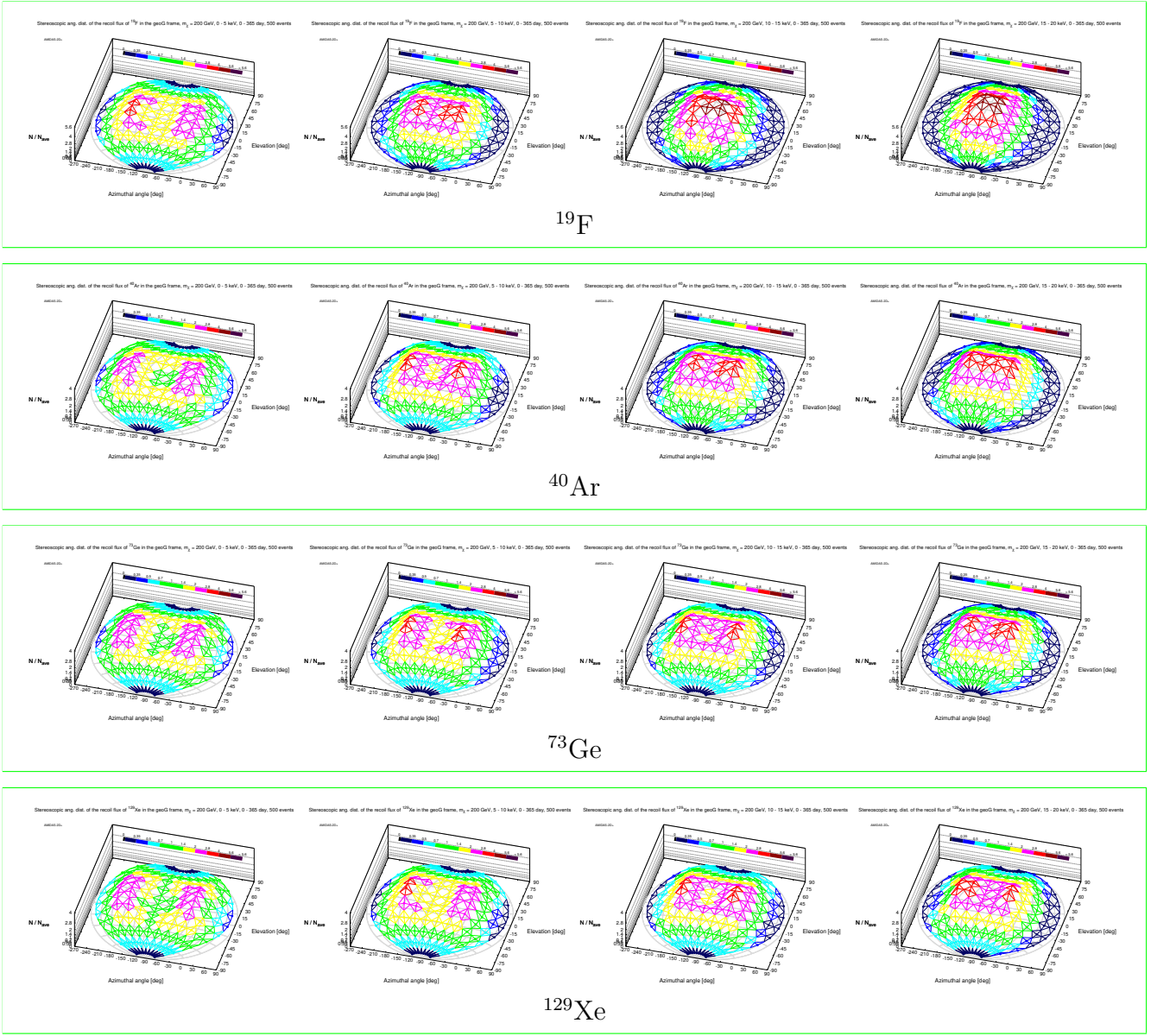
   except that
   the mass of incident halo WIMPs is raised to $\mchi = 200$ GeV.%
   }

 Now
 the stereoscopic angular recoil--energy spectra
 become to a (latitudinally) ``crater--like'' structure:
 the lower the recoil energy,
 the wider the central plain of
 the crater.
 With heavy nuclei
 (e.g.~%
  $\rmXA{I}{127}$,
  $\rmXe$,
  $\rmXA{Cs}{133}$,
  and $\rmW$)
 as our detector materials,
 this crater--like structure could be observed
 in (relatively) high(er) and large(r) energy ranges,
 while
 for light nuclei
 (e.g.~%
  $\rmXA{C}{12}$,
  $\rmF$,
  $\rmXA{S}{32}$,
  $\rmXA{Cl}{35/37}$,
  and $\rmAr$),
 one could observe
 such a crater--like structure
 only in (pretty) low energy ranges.

 Moreover,
 one can also find the ``inner--outer--asymmetry''
 discussed in Ref.~\cite{DMDDD-v_theta}:
 while
 for a WIMP mass of a few tens GeV,
 (the decrease of) the recoil fluxes
 on the inner (east) sky
 are larger (and sharper)
 than those on the outer (west) sky,
 once the WIMP mass is
 as heavy as a few hundreds GeV,
 (the decrease of) the recoil fluxes
 on the outer sky
 could become larger (and sharper).
 Additionally,
 the heavier the target nuclei,
 the higher the upper limit of the recoil energy range,
 in which
 one could observe this ``reverse'' inner--outer--asymmetry.

\section{Summary}

 In this article,
 as an extension of
 our study on
 the angular distribution of
 the recoil flux of
 WIMP--scattered target nuclei,
 we have demonstrated
 the possibility of
 determining the mass of incident halo WIMPs
 by using or combining
 the ``ridge--crater'' structures of
 the angular recoil--energy spectra
 with different target nuclei
 observed in directional direct DM detection experiments.

 Our simulation results show that,
 once the WIMP mass is as light as only a few tens GeV,
 the stereoscopic angular recoil--flux distributions of
 both of light and heavy target nuclei
 would have the (longitudinally) ``ridge--like'' structure:
 the higher the recoil energy,
 the more concentrated the angular recoil--flux distribution
 and (relatively) the higher the crest--line/peak.
 In addition,
 the heavier the target nuclei we use,
 the more obviously and rapidly
 the ridge--like structure of
 the angular recoil--flux distributions
 as well as
 the concentration and the increase of
 the angular recoil--energy spectrum
 would be.

 In contrast,
 once the WIMP mass is as heavy as a few hundreds GeV,
 the angular recoil--flux distributions of
 heavy target nuclei
 would show the (latitudinally) ``crater--like'' structure:
 the lower the recoil energy,
 the wider the central plain of
 the angular recoil--flux distribution,
 whereas
 the distributions of
 light target nuclei
 could have such a crater--like structure
 only in (pretty) low energy range.

 These WIMP--mass and target--nucleus dependent characteristics of
 the angular recoil--energy spectrum
 indicate the possibility of
 (analytic and perhaps model--independent) reconstruction(s) of
 the WIMP mass
 (and perhaps some other WIMP properties)
 by comparing and/or combining
 the angular recoil--energy spectra
 offered by directional direct detection experiments
 with one or several target nuclei.

\subsubsection*{Acknowledgments}

 This work
 was strongly encouraged by
 the ``{\it Researchers working on
 e.g.~exploring the Universe or landing on the Moon
 should not stay here but go abroad.}'' speech.

%
%
%
%

%
%

%
%
%
\end{document}